\documentclass[preprint,12pt]{elsarticle}
\usepackage{amssymb}
\usepackage{amsmath}
\usepackage[numbers]{natbib}
\usepackage{tikz} 
\usepackage{mathtools}
\usetikzlibrary{arrows.meta} 
\usetikzlibrary{decorations.markings}  
\tikzset{midarrow/.style={postaction={decorate,decoration={markings,mark=at position 0.74 with {\arrow{Latex[length=3mm]}}}}}}
\cortext[cor1]{Corresponding author}

\begin{document}
\title{Quasiparticle properties of a single $\Lambda$ impurity in symmetric nuclear matter with a regulated $N\Lambda$ interaction}  

\author[1]{Bahruz Suleymanli}
\ead{bahruz.suleymanli@gmail.com}
\affiliation[1]{organization={Physics Department, Yıldız Technical University},
                city={Istanbul},
                postcode={34220}, 
                state={Esenler},
                country={Türkiye}}
\author[1,2]{Kutsal Bozkurt}
\affiliation[2]{organization={Université Paris-Saclay, CNRS/IN2P3, IJCLab},
                city={Orsay},
                postcode={91405}, 
                country={France}}

\begin{abstract}

We explore the quasiparticle properties of a single $\Lambda$ hyperon propagating through symmetric nuclear matter using the Green's function formalism. The $N\Lambda$ interaction is described by a non-local regulated low-momentum contact potential with a leading-order constant term and a next-to-leading-order derivative correction. The two coupling constants in the ${}^1S_0$ and ${}^3S_1$ channels are fixed by matching the vacuum on-shell $T$ matrix to the scattering length and effective range obtained from modern next-to-next-to-leading-order chiral effective field theory. Using this effective interaction, we calculate the retarded $\Lambda$ self-energy from the in-medium $N\Lambda$ ladder $T$ matrix, which sums repeated $N\Lambda$ scattering in the nucleonic medium. At saturation density, the zero-momentum quasiparticle pole is found at $E_{\rm qp}(0,\rho_{\rm sat})=-29.55~{\rm MeV}$, in good agreement with the empirical depth of the single $\Lambda$ potential in nuclear matter. The self-energy decomposition gives a static Born contribution $\Sigma_\Lambda^{\rm Born}(0)=-26.36~{\rm MeV}$ and a dynamical correlation contribution ${\rm Re}\,\Sigma_\Lambda^{\rm corr,R}(0,E_{\rm qp})=-3.19~{\rm MeV}$, showing that repeated in-medium $N\Lambda$ scattering is needed to reproduce the empirical binding scale. The quasiparticle remains narrow and well defined, with a large residue $Z(0)=0.98$, a small damping width $\Gamma(0)=0.023~{\rm MeV}$, and a sharp spectral peak near the quasiparticle energy. At finite momentum, the $\Lambda$ quasiparticle becomes less bound, with $E_{\rm qp}(k,\rho_{\rm sat})$ increasing from $-29.55~{\rm MeV}$ at $k=0$ to $-6.49~{\rm MeV}$ at $k=1~{\rm fm}^{-1}$, while the residue and width change only weakly. A low-momentum fit gives $m_\Lambda^*/m_\Lambda=0.747$, consistent with the range obtained in Brueckner calculations with Nijmegen hyperon--nucleon potentials. These results show that a compact effective-range-constrained $N\Lambda$ interaction, combined with an in-medium ladder self-energy, gives a realistic description of the $\Lambda$ binding, spectral strength, damping width, and effective mass in symmetric nuclear matter.

\end{abstract}

\maketitle

\section{Introduction}\label{sec:1}
The extension of nuclear many-body physics into the strangeness sector provides
a unique probe of baryonic interactions in a nuclear medium, from
single $\Lambda$ hypernuclei to more complex strange hadronic systems
~\cite{Chrien1989,Gal2016,Miwa2025}.  In view of the scarcity of experimental
data, one of the most important and challenging baryonic interactions is the
interaction between nucleons ($N$) and $\Lambda$ hyperons.  Even in the
single $\Lambda$ sector, the available data are limited, while the situation is
even more restrictive for double-$\Lambda$ systems, which makes a quantitative
understanding of the $N\Lambda$ interaction difficult~\cite{Vidana2018,He2025}.
Recent femtoscopic measurements have opened a
complementary route to this problem, showing that momentum correlations of
hadron pairs produced at the LHC can provide detailed information on the
short-range part of the $N\Lambda$ interaction~\cite{ALICE2020}.
Nevertheless, $\Lambda$ hyperons have a special advantage as probes of the
nuclear interior~\cite{Nakamura2026}.  Since the $\Lambda$ hyperon is a different
baryon from the nucleons, the occupied proton ($p$) and neutron ($n$) states do not
Pauli-block its motion.  As a result, a $\Lambda$ hyperon can penetrate deeply
into the nucleus and behave as an impurity probe of the nuclear medium.  This
makes the single $\Lambda$ impurity limit a natural benchmark for microscopic
many-body theory, because the $\Lambda$ self-energy in symmetric nuclear
matter directly connects the underlying $N\Lambda$ interaction to observable
quantities such as the $\Lambda$ binding energy, spectral strength, damping
width, and effective mass.  This limit is also directly connected to
single $\Lambda$ hypernuclear spectroscopy, where $\Lambda$ separation
energies, shell states, weak spin-orbit splittings, and core shrinkage effects
show that the $\Lambda$ behaves as a well-defined quasiparticle in the nuclear
environment~\cite{Ajimura2001,Tanida2001,Hotchi2001,Garibaldi2019}.

The role of $\Lambda$ hyperons becomes even more important in dense matter.
In beta-equilibrated neutron-star matter, the neutron chemical potential
increases with density, and it may become energetically favorable to replace
some high-momentum neutrons by strange baryons, with the $\Lambda$ usually
being the first hyperon to appear \cite{Weber2007}.  Brueckner-Hartree-Fock calculations showed
that the inclusion of $\Lambda$ and other hyperons can strongly modify the
composition of neutron-star matter and soften the equation of state
~\cite{Schulze1998,Vidana2000}.  This softening reduces the maximum neutron-star
mass and became especially problematic after the observation of neutron stars
with masses close to two solar masses~\cite{Demorest2010,Antoniadis2013,
Cromartie2020}.  This tension is the so-called hyperon puzzle: hyperons are
expected on energetic grounds, but their appearance must be reconciled with a
sufficiently stiff equation of state~\cite{Vidana2018,Burgio2021}.
Several works have shown that the solution is highly sensitive to the
in-medium $\Lambda N$ interaction, possible $\Lambda NN$ three-body forces, and
the density dependence of the $\Lambda$ single-particle potential
~\cite{Lonardoni2015,Haidenbauer2017,Providencia2019}.  Therefore, even before
constructing a full hyperonic equation of state, it is essential to establish
how a single $\Lambda$ quasiparticle is generated and renormalized by the
nucleonic medium.

Green's-function techniques have become standard tools for treating correlated
nuclear systems, from self-consistent calculations of spectral functions and
quasiparticle lifetimes in nuclear matter to recent impurity-induced resonance
problems in low-energy nuclear scattering~\cite{Dickhoff2004,Rios2012,Soma2013,Soma2020,Suleymanli2026}.
Using this technique, studies of the propagation of a $\Lambda$ hyperon in
nuclear matter have shown that short-range $N\Lambda$ correlations shift part
of the single-particle strength away from the quasiparticle peak, although the
$\Lambda$ remains less correlated than a nucleon~\cite{Robertson2004}.
Microscopic finite-nucleus calculations had already
emphasized that the $\Lambda$ self-energy is nonlocal and energy dependent,
and that correlations beyond the Hartree--Fock approximation affect both bound
and scattering $\Lambda$ states~\cite{Hjorth1996}.
Related calculations
of finite hypernuclei constructed the $\Lambda$ spectral function from the
$\Lambda$ self-energy and found relatively large quasiparticle strengths,
supporting the picture that the $\Lambda$ largely preserves its identity inside
the nuclear medium~\cite{Vidana2017}.  More recent studies based on chiral
$YN$ interactions have further used the $\Lambda$ self-energy as the bridge
between modern $N\Lambda$ forces and the single-particle spectra of
$\Lambda$ hypernuclei~\cite{Haidenbauer2020}. Beyond these studies, Green's-function analyses that directly connect the $N\Lambda$ interaction to the $\Lambda$ quasiparticle spectrum in the clean
single-impurity limit remain limited.
 
A practical way to connect the poorly constrained $N\Lambda$ force with
many-body observables is to use a low-momentum short-range representation, in
which the unknown short-distance dynamics is absorbed into contact terms,
the leading momentum dependence is included through derivative corrections,
and a finite momentum-space regulator keeps the interaction within its
low-energy domain~\cite{Haidenbauer2020,Haidenbauer2013,Haidenbauer2023,
Ren2020,Petschauer2020,Song2022}.
Motivated by this picture, we study the correlated many-body problem
in the strangeness sector, where a single $\Lambda$ hyperon embedded in symmetric
nuclear matter.  The interaction is taken in the spin-averaged $S$-wave
$N\Lambda$ channel appropriate for spin-saturated matter, and repeated
in-medium $N\Lambda$ scattering is summed through a ladder $T$ matrix.  This
allows us to determine the retarded $\Lambda$ self-energy, spectral function,
quasiparticle energy, damping width, and effective mass, and to quantify how
much of the empirical $\Lambda$ binding originates from the static Born term
and how much from nonperturbative ladder correlations.

The present manuscript is organized as follows.  In Sec.~\ref{sec:2}, we introduce the single $\Lambda$ impurity formalism, define the regulated $N\Lambda$ interaction, and derive the retarded self-energy from the in-medium ladder $T$ matrix.  In Sec.~\ref{sec:3}, we present the numerical results for the zero- and finite-momentum $\Lambda$ quasiparticle properties, including the self-energy, spectral function, damping width, and effective mass.  Conclusions are given in Sec.~\ref{sec:conc}. \ref{sec:A1} gives the mapping procedure used to determine the $N\Lambda$ contact couplings from the scattering length and effective range.

\section{Green's Function Formalism for a Single $\Lambda$ Impurity}\label{sec:2}

We study the propagation of a single $\Lambda$ hyperon in a uniform system of symmetric nuclear matter. In this symmetric medium, the total nucleonic density $\rho_N$ consists of equal parts neutrons and protons ($\rho_n = \rho_p = \rho_N / 2$). In the limit of a single embedded hyperon, the $\Lambda$ density is negligible ($\rho_\Lambda = 0$). Consequently, the nucleons form a dense, interacting background medium, while the $\Lambda$ hyperon is treated as a distinct impurity quasiparticle probing this nucleonic environment.
The corresponding Hamiltonian is given by
\begin{equation}
H=H_0+H_{N\Lambda},
\label{eq:H_full_single_lambda}
\end{equation}
where $H_0$ is the one-body Hamiltonian and $H_{N\Lambda}$ describes the
interaction between the impurity $\Lambda$ and the surrounding nucleons. In
second quantization,
\begin{equation}
H_0
=
\sum_{\sigma}
\int\!\frac{d^3k}{(2\pi)^3}
\left(
\frac{\hbar^2 k^2}{2m_N}-\mu_N
\right)
a^\dagger_{N\sigma}(\mathbf k)a_{N\sigma}(\mathbf k)
+
\sum_{\sigma}
\int\!\frac{d^3k}{(2\pi)^3}
\frac{\hbar^2 k^2}{2m_\Lambda}
a^\dagger_{\Lambda\sigma}(\mathbf k)a_{\Lambda\sigma}(\mathbf k).
\label{eq:H0_single_lambda}
\end{equation}
Here, $a_{B\sigma}(\mathbf k)$ and $a^\dagger_{B\sigma}(\mathbf k)$ are the annihilation and creation operators for the baryon $B=N,\Lambda$, $\sigma=\uparrow,\downarrow$ is the spin projection, and $\mu_N$ is the chemical potential of the nucleons. Since we work in the single $\Lambda$ limit, no $\Lambda$ chemical
potential is needed for the medium occupation.
The interaction Hamiltonian in center-of-mass coordinates can be written as follows:
\begin{eqnarray}
H_{N\Lambda}
&=&
\sum_{\sigma_1\sigma_2\sigma_3\sigma_4}
\int
\frac{d^3K}{(2\pi)^3}
\frac{d^3p}{(2\pi)^3}
\frac{d^3p'}{(2\pi)^3}
\,
a^\dagger_{\Lambda\sigma_3}\!\left(\frac{\mathbf K}{2}+\mathbf p'\right)
a^\dagger_{N\sigma_4}\!\left(\frac{\mathbf K}{2}-\mathbf p'\right)
\nonumber\\
&&\times
V^{\sigma_3\sigma_4;\sigma_1\sigma_2}_{N\Lambda}
(\mathbf p',\mathbf p)
a_{N\sigma_2}\!\left(\frac{\mathbf K}{2}-\mathbf p\right)
a_{\Lambda\sigma_1}\!\left(\frac{\mathbf K}{2}+\mathbf p\right).
\label{eq:H_NL}
\end{eqnarray}
Here $\mathbf K=\mathbf k_N+\mathbf k_\Lambda$ is the total momentum of the
$N\Lambda$ pair, while
$\mathbf p=(\mathbf k_\Lambda-\mathbf k_N)/2$ and
$\mathbf p'=(\mathbf k'_\Lambda-\mathbf k'_N)/2$ are the initial and final
relative momenta, respectively.
For the low-energy $N\Lambda$ interaction we use a regulated contact form with
a leading term and the first derivative correction~\cite{Haidenbauer2023,Lorenzi2023},
\begin{equation}
V_{N\Lambda}^{S}(\mathbf p',\mathbf p)
=
\left[
g_{0,N\Lambda}^{S}
+
g_{2,N\Lambda}^{S}
\left(p'^2+p^2\right)
\right]
\exp\left[-\frac{p'^2+p^2}{\Lambda_{\rm cut}^2}\right],
\label{eq:V_NL_regulated}
\end{equation}
where $S=0,1$ denotes the spin-singlet and spin-triplet channels. The leading-order (LO) constant $g_{0,N\Lambda}^{S}$ and the next-to-leading-order
(NLO) derivative constant $g_{2,N\Lambda}^{S}$ are fixed from the corresponding
$S$-wave scattering length ($a_{N\Lambda}^S$) and effective range ($r_{N\Lambda}^S$) by solving the vacuum two-body
Lippmann--Schwinger equation and matching the on-shell $T$ matrix to the
effective-range expansion, with the details of this procedure given in
\ref{sec:A1}. The Gaussian cutoff
$\Lambda_{\rm cut}$ suppresses high-momentum components and defines the
low-momentum domain of the effective interaction. In this work, calculations are performed at $\Lambda_{\rm cut}=500~{\rm MeV}$, where the application of the $a_{N\Lambda}^S$ and $r_{N\Lambda}^S$ values from Ref.~\cite{Haidenbauer2023} to our formalism in \ref{sec:A1} yields the values for $g_{0,N\Lambda}^{S}$ and $g_{2,N\Lambda}^{S}$ listed in Table~\ref{tab:scatt_input}. For spin-saturated symmetric nuclear matter, the $N\Lambda$ spin states are
populated according to their degeneracies. The singlet channel has weight
$1/4$, while the triplet channel has weight $3/4$~\cite{Sakurai2021}. We therefore use
\begin{equation}
V_{N\Lambda}^{\rm av}
=
\frac{1}{4}V_{N\Lambda}^{\,{}^1S_0}
+
\frac{3}{4}V_{N\Lambda}^{\,{}^3S_1}.
\label{eq:V_NL_average}
\end{equation}
The corresponding averaged coupling constants are $g_{0,N\Lambda}^{\rm av}
=
\frac14 g_{0,N\Lambda}^{\,{}^1S_0}
+
\frac34 g_{0,N\Lambda}^{\,{}^3S_1}
=
-1.03~{\rm fm}^2,$
and
$g_{2,N\Lambda}^{\rm av}
=
\frac14 g_{2,N\Lambda}^{\,{}^1S_0}
+
\frac34 g_{2,N\Lambda}^{\,{}^3S_1}
=
0.34~{\rm fm}^4.$

\begin{table}[t]
\centering
\caption{
Calculated leading-order and next-to-leading-order $N\Lambda$ coupling constants, $g_{0,N\Lambda}^{S}$ and $g_{2,N\Lambda}^{S}$, obtained at a momentum cutoff of $\Lambda_{\rm cut}=500~{\rm MeV}$. The calculations utilize the scattering length $a_{N\Lambda}^S$ and effective range $r_{N\Lambda}^S$ parameters adopted from Ref.~\cite{Haidenbauer2023}.}
\label{tab:scatt_input}
\begin{tabular}{c c c c c}
\hline\hline
 Partial wave & $a_{N\Lambda}^S$ (fm) & $r_{N\Lambda}^S$ (fm) &
$g_{0,N\Lambda}^{S}$ (fm$^2$) &
$g_{2,N\Lambda}^{S}$ (fm$^4$)\\
\hline
 ${}^1S_0$ & $-2.80$ & $2.87$ & $-1.15$ & $0.50$ \\
 ${}^3S_1$ & $-1.59$ & $3.10$ & $-1.06$ & $0.28$ \\
\hline\hline
\end{tabular}
\end{table}

Based on the Hamiltonian given in Eq.~(\ref{eq:H_full_single_lambda}), we can express the retarded Green's function via the Dyson equation as follows:
\begin{equation}
G_\Lambda^{R}(\mathbf k,\omega)
=
\frac{1}
{\omega-\varepsilon_\Lambda(k)
-\Sigma_\Lambda^{R}(\mathbf k,\omega)},
\label{eq:Dyson_Lambda_R}
\end{equation}
where $\varepsilon_\Lambda(k)=\hbar^2 k^2/\left(2m_\Lambda\right)$.
The self-energy $\Sigma_\Lambda^R$ contains the effect of the nucleonic medium
on the propagation of the impurity. This Green's-function formulation directly
connects the self-energy to the quasiparticle energy, spectral function,
residue, damping width, and effective mass~\cite{Dickhoff2004,Rios2012,Fetter2003}.
Beyond the static Hartree--Fock approximation, we include repeated in-medium
$N\Lambda$ scattering through the retarded in-medium $T$ matrix. In this
formulation the self-energy is evaluated directly from the full ladder
$T$ matrix \cite{Fetter2003}. 
For the calculation of a single $\Lambda$ impurity in symmetric nuclear matter, the retarded $\Lambda$ self-energy can be determined as follows:
\begin{equation}
\Sigma_\Lambda^{R}(\mathbf k,\omega)
=
2\sum_{N=n,p}
\int\!\frac{d^3k'}{(2\pi)^3}
\, n_N(\mathbf k')\,
T_{N\Lambda}^{R}
\!\left(
q_{\rm rel},q_{\rm rel};
\Omega,\mathbf K
\right),
\label{eq:Sigma_T_single_Lambda}
\end{equation}
where $q_{\rm rel}=\frac12|\mathbf k-\mathbf k'|, \Omega=\omega+\xi_N(k')$, and
$n_N(\mathbf k')=\theta(k_{F,N}-k')$ is the zero-temperature Fermi occupation
number of nucleons in the symmetric nuclear medium.
The nucleon single-particle energy relative to the Fermi surface is
$\xi_N(k)=\frac{\hbar^2k^2}{2m_N}-\mu_N$.
For symmetric nuclear matter,
$k_{F,n}=k_{F,p}=(3\pi^2\rho_N/2)^{1/3}$. 
In the present calculation, the $\Lambda$ is an impurity and is not Pauli
occupied. Therefore the in-medium two-particle propagator contains only the
nucleon Pauli blocking factor,
\begin{equation}
\Pi_{N\Lambda}^R(\mathbf K,\mathbf q;\Omega)
=
\frac{1-n_N(k_N)}
{\Omega-\xi_N(k_N)-\varepsilon_\Lambda(k_\Lambda)+i\eta}.
\label{eq:Pi_NL_R}
\end{equation}
Here $k_N$ and $k_\Lambda$ are the nucleon and $\Lambda$ momenta in the
intermediate $N\Lambda$ pair. In the numerical calculation we use a small
positive broadening parameter $\eta$.
Because the interaction in Eq.~(\ref{eq:V_NL_regulated}) is separable, the
in-medium ladder equation reduces to a finite matrix inversion. We define
\begin{equation}
f_0(p)=e^{-p^2/\Lambda_{\rm cut}^2},
\qquad
f_2(p)=p^2 e^{-p^2/\Lambda_{\rm cut}^2},
\label{eq:f0_f2}
\end{equation}
and write the spin-averaged interaction as
\begin{equation}
V_{N\Lambda}^{\rm av}(p',p)
=
\sum_{a,b=0,2}
f_a(p')\,\lambda_{ab}\,f_b(p),
\qquad
\lambda
=
\begin{pmatrix}
g_0 & g_2\\
g_2 & 0
\end{pmatrix}.
\label{eq:V_separable}
\end{equation}
Here $g_0\equiv g_{0,N\Lambda}^{\rm av}$ and
$g_2\equiv g_{2,N\Lambda}^{\rm av}$. The retarded in-medium $T$ matrix is then
written in the same basis,
\begin{equation}
T_{N\Lambda}^{R}(p',p;\Omega,K)
=
\sum_{a,b=0,2}
f_a(p')\,
\tau_{ab}^{R}(\Omega,K)\,
f_b(p).
\label{eq:T_separable}
\end{equation}
The matrix $\tau^R$ is
\begin{equation}
\tau^R(\Omega,K)
=
\left[
\lambda^{-1}
-
J^R(\Omega,K)
\right]^{-1},
\label{eq:tau_matrix_1}
\end{equation}
with
\begin{equation}
J_{ab}^R(\Omega,K)
=
\int\!\frac{d^3q}{(2\pi)^3}
f_a(q)\,
\Pi_{N\Lambda}^R(\mathbf K,\mathbf q;\Omega)\,
f_b(q).
\label{eq:J_matrix_1}
\end{equation}
Equations~(\ref{eq:T_separable})--(\ref{eq:J_matrix_1}) are the in-medium
ladder resummation in the separable basis. Ladder and $T$-matrix resummations of this type are standard tools
for treating repeated two-body scattering in nuclear many-body theory
~\cite{Dickhoff2004,Rios2012,Fetter2003}. By evaluating the retarded $\Lambda$ self-energy, we can determine the quasiparticle energy,
\begin{equation}
\omega
-
\frac{\hbar^2 k^2}{2m_\Lambda}
-
{\rm Re}\,\Sigma_\Lambda^{R}(k,\omega)
=
0,
\label{eq:qp_finite_k}
\end{equation}
the quasiparticle residue,
\begin{equation}
Z(k)
=
\left[
1-
\left.
\frac{\partial}{\partial\omega}
{\rm Re}\,\Sigma_\Lambda^{R}(k,\omega)
\right|_{\omega=E_{\rm qp}(k)}
\right]^{-1},
\label{eq:Z_factor}
\end{equation}
and the damping width,
\begin{equation}
\Gamma(k)
=
-2Z(k)\,
{\rm Im}\,\Sigma_\Lambda^{R}
\!\left(k,E_{\rm qp}(k)\right),
\label{eq:Gamma_width}
\end{equation}
using the following relations. Furthermore, the corresponding spectral function can be extracted from the Green's function in Eq.~(\ref{eq:Dyson_Lambda_R}) as follows:
\begin{equation}
A_\Lambda(k,\omega)
=
-\frac{1}{\pi}
{\rm Im}\,
G_\Lambda^R(k,\omega).
\label{eq:spectral_function}
\end{equation}
Finally, the effective mass is extracted from the low-momentum quasiparticle
dispersion. We write
\begin{equation}
E_{\rm qp}(k, \rho_N)-E_{\rm qp}(0, \rho_N)
\simeq
c k^2,
\label{eq:low_k_fit}
\end{equation}
and define
\begin{equation}
\frac{m_\Lambda^*}{m_\Lambda}
=
\frac{\hbar^2/(2m_\Lambda)}{c}.
\label{eq:mstar_fit}
\end{equation}
This definition includes both the explicit momentum dependence of the
self-energy and its energy dependence through the quasiparticle equation.

\section{Results}\label{sec:3}

In this section, we present the results obtained using the formalism developed in Section~\ref{sec:2}. Focusing on symmetric nuclear matter at saturation density ($\rho_n = \rho_p = \rho_{\rm sat} / 2$), we discuss the zero-momentum case in Section~\ref{sec:3.1} and present finite-momentum calculations in Section~\ref{sec:3.2}.

\subsection{Zero-momentum $\Lambda$ quasiparticle}\label{sec:3.1}

We start our numerical analysis by determining the quasiparticle properties at
zero momentum. This case is the natural benchmark for the calculation,
because analyses of $(\pi^+,K^+)$ and $(K^-,\pi^-)$ reactions, together with
systematics of $\Lambda$ binding energies, show that the depth of the
single $\Lambda$ potential in nuclear matter is approximately~$U_\Lambda(0,\rho_{\rm sat})\simeq -(28\text{--}30)$ MeV~\cite{Schulze1998,Millener1988,Motoba1995,Friedman2023}. In what follows, we examine how the calculated $\Lambda$ self-energy generates
this binding scale and how it determines the quasiparticle residue, damping
width, and spectral function at $k=0$.

\subsubsection{Binding energy from the quasiparticle pole}
\begin{figure}
    \centering
    \includegraphics[width=0.8\textwidth]{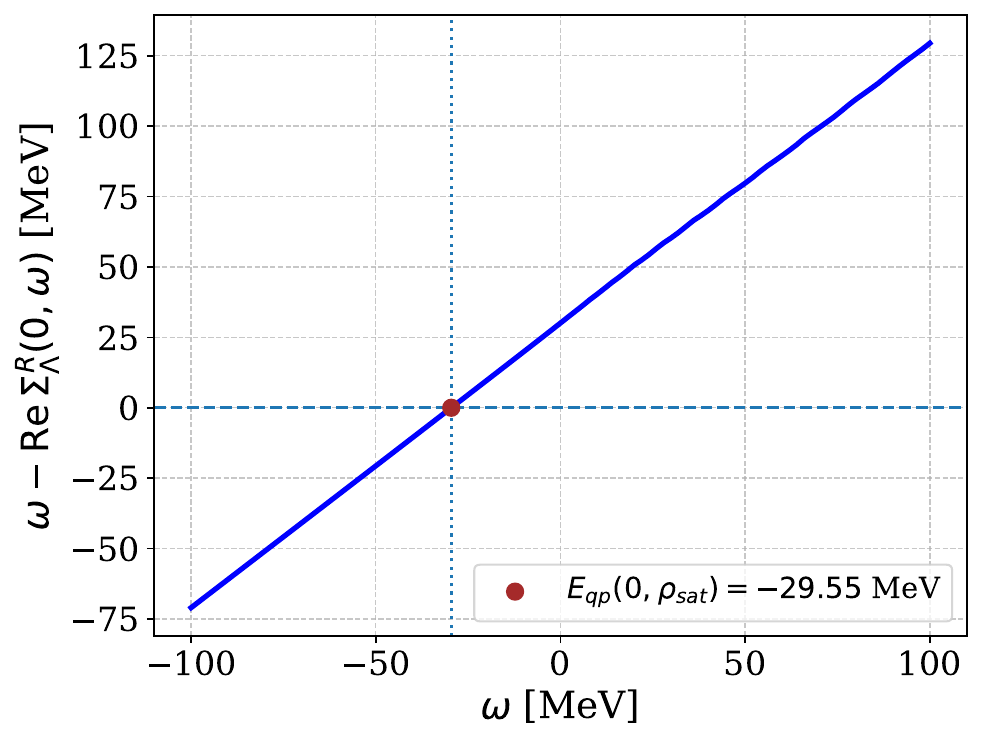}
    \caption{Determination of the zero-momentum $\Lambda$ quasiparticle energy
from Eq.~\eqref{eq:qp_k0}. The root of
$\omega-{\rm Re}\,\Sigma_\Lambda^R(0,\omega)$ gives
$E_{\rm qp}(0,\rho_{\rm sat})=-29.55~{\rm MeV}$.}
    \label{fig:Eqp}
\end{figure}

At $k=0$, Eq.~(\ref{eq:qp_finite_k}) reduces to the following form:
\begin{equation}
\omega
-
{\rm Re}\,\Sigma_\Lambda^{R}(0,\omega)
=
0.
\label{eq:qp_k0}
\end{equation}
Using the self-energy in Eq.~(\ref{eq:Sigma_T_single_Lambda}), the solution of
the quasiparticle equation is shown graphically in Fig.~\ref{fig:Eqp}. As can
be seen from the figure, the value of $\omega$ satisfying Eq.~(\ref{eq:qp_k0}),
namely the zero-momentum $\Lambda$ quasiparticle energy, is
\begin{equation}\label{eq:Eqp_k0_main}
E_{\rm qp}(0,\rho_{\rm sat})=-29.55~{\rm MeV}.
\end{equation}
This value is in good agreement with the empirical depth of the $\Lambda$
single-particle potential in nuclear matter.
The result in Fig.~\ref{fig:Eqp} therefore demonstrates that repeated
in-medium $N\Lambda$ scattering generates a realistic attractive
$\Lambda$ quasiparticle potential. In contrast to a purely static mean-field
picture, the present calculation includes the full ladder $T$ matrix in the
self-energy. Thus the position of the quasiparticle pole contains both the
Born contribution and the dynamical correlation contribution associated with
successive $N\Lambda$ scatterings in the nuclear medium. This provides a
microscopic Green's-function interpretation of the empirical
$\Lambda$ potential depth: the $\Lambda$ impurity becomes a well-defined
quasiparticle whose binding is generated by the in-medium $N\Lambda$ ladder
self-energy.

\subsubsection{Static and dynamical contributions to the $\Lambda$ self-energy}\label{sec:3.1.2}
\begin{figure}
    \centering
    \includegraphics[width=0.75\textwidth]{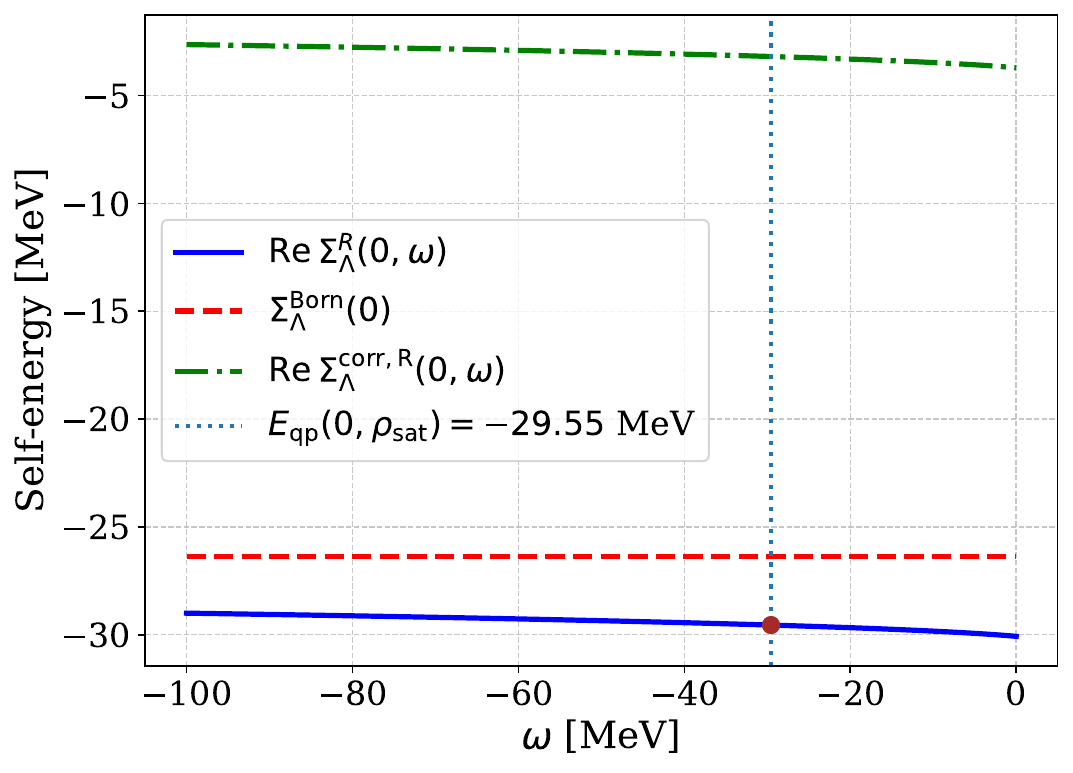}
    \caption{Energy dependence of the zero-momentum $\Lambda$ self-energy.  The blue solid line shows
    ${\rm Re}\,\Sigma_\Lambda^{R}(0,\omega)$, the red dashed line shows the static
    Born contribution $\Sigma_\Lambda^{\rm Born}(0)$, and the green dash-dotted
    line shows the correlation contribution
    ${\rm Re}\,\Sigma_\Lambda^{\rm corr,R}(0,\omega)$.
    }
    \label{fig:Sigma_Born_cor}
\end{figure}

To understand the origin of the binding, we decompose the full $T$-matrix
self-energy into its Born and correlation parts,
\begin{equation}
\Sigma_\Lambda^{R}(0,\omega)
=
\Sigma_\Lambda^{\rm Born}(0)
+
\Sigma_\Lambda^{\rm corr,R}(0,\omega),
\label{eq:Sigma_Born_corr}
\end{equation}
where
\begin{equation}
\Sigma_\Lambda^{\rm Born}(0)
=
2\sum_{N=n,p}
\int\!\frac{d^3k'}{(2\pi)^3}\,
n_N(\mathbf k')\,
V_{N\Lambda}^{\rm av}
\left(q_{\rm rel},q_{\rm rel}\right).
\label{eq:Sigma_Born_k0_general}
\end{equation}
Here $\Sigma_\Lambda^{\rm Born}$ is equivalent to the Hartree--Fock
contribution, while $\Sigma_\Lambda^{\rm corr,R}$ contains the repeated
in-medium $N\Lambda$ scatterings generated by the ladder $T$ matrix.
Fig.~\ref{fig:Sigma_Born_cor} shows the energy dependence of
${\rm Re}\,\Sigma_\Lambda^{R}(0,\omega)$ obtained from
Eq.~(\ref{eq:Sigma_Born_corr}), together with the Born contribution
$\Sigma_\Lambda^{\rm Born}(0)$ from
Eq.~(\ref{eq:Sigma_Born_k0_general}) and the correlation contribution
${\rm Re}\,\Sigma_\Lambda^{\rm corr,R}(0,\omega)
=
{\rm Re}\,\Sigma_\Lambda^{R}(0,\omega)
-
\Sigma_\Lambda^{\rm Born}(0)$, in the negative-energy region.  The results show
that the largest part of ${\rm Re}\,\Sigma_\Lambda^{R}(0,\omega)$ comes from
the Born contribution.  Since this term is static, it does not depend on
$\omega$ and takes the value $\Sigma_\Lambda^{\rm Born}(0)=-26.36~{\rm MeV}.$
However, this value alone is still not sufficient to reproduce the empirical
depth of the single $\Lambda$ potential.  The remaining attraction is supplied
by the correlation contribution generated by repeated in-medium $N\Lambda$
scattering.  At the quasiparticle pole,
$E_{\rm qp}(0,\rho_{\rm sat})=-29.55~{\rm MeV}$, we find
\begin{equation}
{\rm Re}\,\Sigma_\Lambda^{\rm corr,R}(0,E_{\rm qp})=-3.19~{\rm MeV}.
\end{equation}
Thus, although the Born term provides the dominant static attraction, the
correlation part is essential for reaching the empirical binding scale.  This
shows that the $\Lambda$ potential depth cannot be understood only as a
first-order mean-field effect.  Even in the single-impurity limit, repeated
$N\Lambda$ scattering gives a non-negligible dynamical correction to the
$\Lambda$ self-energy and is required for a realistic quasiparticle energy.

\subsubsection{Quasiparticle residue and damping width}\label{sec:3.1.3}

The determination of ${\rm Re}\,\Sigma_\Lambda^{R}(0,\omega)$ in
Fig.~\ref{fig:Sigma_Born_cor} allows us to calculate the quasiparticle residue
by using Eq.~(\ref{eq:Z_factor}).  The quasiparticle residue measures the
strength of the single-particle pole in the full Green's function~(\ref{eq:Dyson_Lambda_R}).  
In other words, it tells us how much of the spectral strength remains concentrated in
the well-defined quasiparticle state rather than being shifted to the
correlated background.  Our calculation gives
\begin{equation}
Z(0)=0.98 .
\end{equation}
This value is close to unity and therefore shows that the zero-momentum
$\Lambda$ remains a very well-defined quasiparticle in symmetric nuclear
matter.  This result is consistent with previous Green's-function and
spectral-function studies, where the $\Lambda$ was found to be less strongly
correlated than a nucleon and to preserve a large quasiparticle strength in the
nuclear medium~\cite{Dickhoff2004,Robertson2004,Vidana2017}.

\begin{figure}
    \centering
    \includegraphics[width=0.8\textwidth]{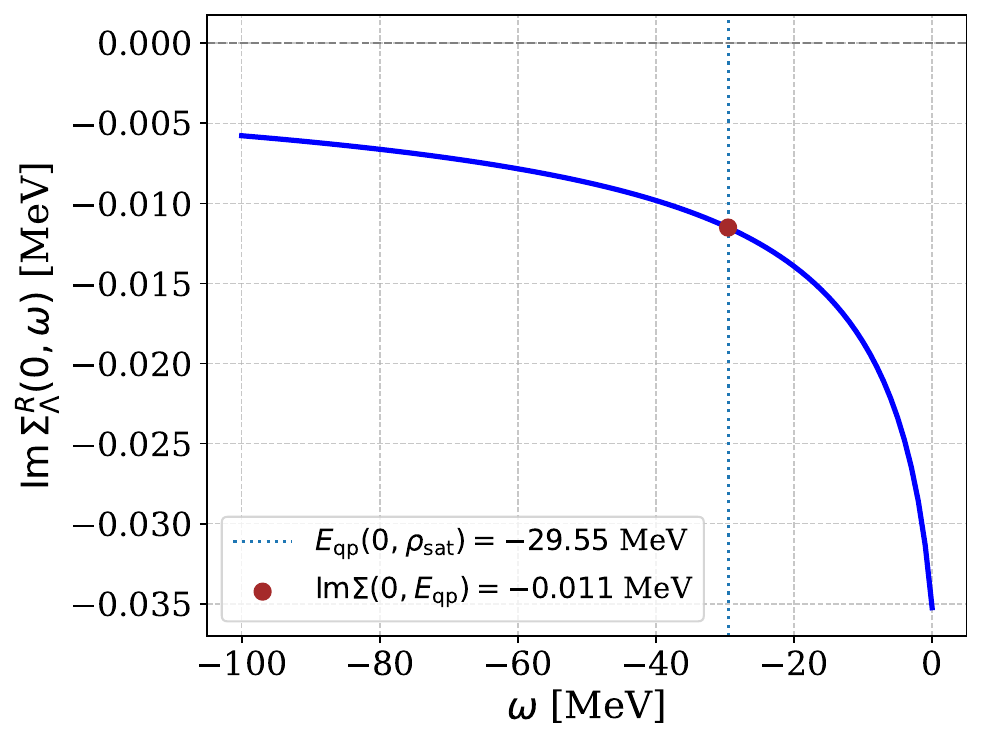}
    \caption{Imaginary part of the zero-momentum retarded $\Lambda$ self-energy as a function of $\omega$. 
    }
    \label{fig:ImSigma_k0}
\end{figure}

The imaginary part of the retarded self-energy, calculated from
Eq.~(\ref{eq:Sigma_T_single_Lambda}), is shown in Fig.~\ref{fig:ImSigma_k0}.
As can be seen, ${\rm Im}\,\Sigma_\Lambda^{R}(0,\omega)$ is negative in the
whole negative-energy region and changes from about $-0.006$ MeV to
$-0.035$ MeV.  Its magnitude increases as $\omega$ becomes larger.  At the
quasiparticle pole, where $E_{\rm qp}$ is given by Eq.~(\ref{eq:Eqp_k0_main}),
we find
\begin{equation}
{\rm Im}\,\Sigma_\Lambda^{R}(0,E_{\rm qp})=-0.011~{\rm MeV}.
\end{equation}
The small value of the imaginary part means that the decay probability of the
$\Lambda$ quasiparticle into correlated many-body states is weak.  Therefore,
the quasiparticle pole is not strongly broadened by in-medium $N\Lambda$
scattering.

Using Eq.~(\ref{eq:Gamma_width}) together with $Z(0)=0.98$ and
${\rm Im}\,\Sigma_\Lambda^{R}(0,E_{\rm qp})=-0.011$ MeV, we obtain
\begin{equation}
\Gamma(0)=0.023~{\rm MeV}.
\end{equation}
This very small damping width confirms that the $\Lambda$ excitation at
$k=0$ is narrow and long lived.  Thus, the real part of the self-energy gives
a realistic attractive quasiparticle energy, while the imaginary part shows
that this quasiparticle remains sharply defined in the nuclear medium.

\subsection{Finite momentum $\Lambda$ quasiparticle}\label{sec:3.2}

In this section, we determine how the $\Lambda$ quasiparticle properties change
at finite momentum for symmetric nuclear matter, where
$\rho_n=\rho_p=\rho_{\rm sat}/2$.  Let us note that extending the calculation to finite
momenta also allows us to determine the effective mass of the $\Lambda$ hyperon
from Eqs.~(\ref{eq:low_k_fit}) and (\ref{eq:mstar_fit}).  Before doing this, we
first analyze how the quasiparticle properties calculated in
Sec.~\ref{sec:3.1} change with $k$.

\begin{figure}
    \centering
    \includegraphics[width=0.95\textwidth]{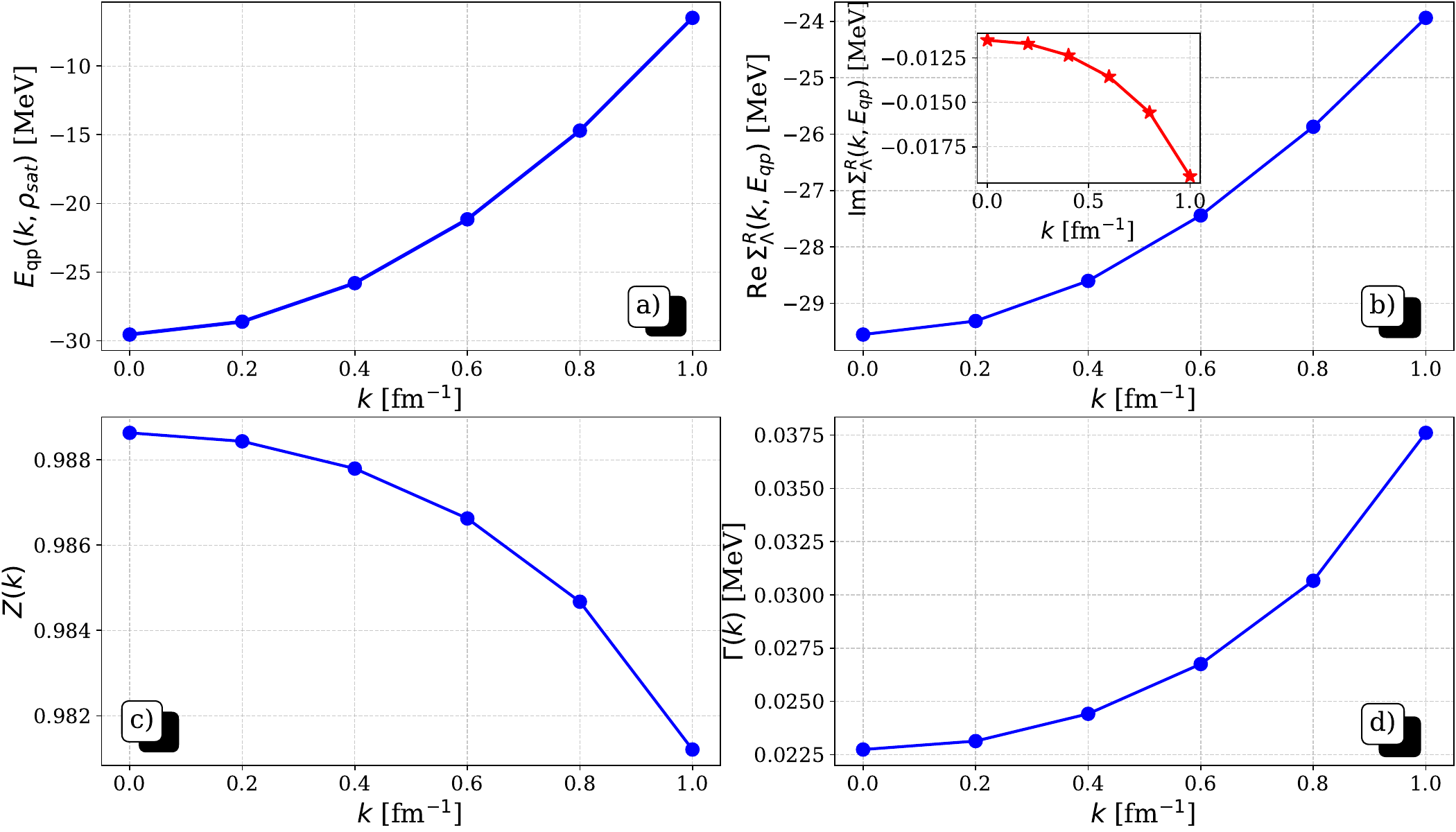}
    \caption{Momentum dependence of the $\Lambda$ quasiparticle properties.  Panel (a) shows the quasiparticle energy
$E_{\rm qp}(k, \rho_{\rm sat})$, panel (b) shows
${\rm Re}\,\Sigma_\Lambda^{R}(k,E_{\rm qp})$, with the inset showing
${\rm Im}\,\Sigma_\Lambda^{R}(k,E_{\rm qp})$, panel (c) shows the
quasiparticle residue $Z(k)$, and panel (d) shows the damping width
$\Gamma(k)$.
    }
    \label{fig:quasi_k}
\end{figure}

Figure~\ref{fig:quasi_k}a shows the quasiparticle energy at finite momenta,
obtained by solving Eq.~(\ref{eq:qp_finite_k}).  As can be seen, when $k$
increases, $E_{\rm qp}(k,\rho_{\rm sat})$ also increases.  Compared with the
self-energy, quasiparticle residue, and damping width, the quasiparticle energy
shows the strongest momentum dependence.  While at $k=0$ it takes the value
given in Eq.~(\ref{eq:Eqp_k0_main}), at $k=1~{\rm fm}^{-1}$ it increases to
$-6.49~{\rm MeV}$.  This behavior shows that the $\Lambda$
quasiparticle becomes less bound as its momentum increases.  In other words,
the attractive in-medium $N\Lambda$ self-energy is not strong enough to keep
the moving $\Lambda$ at the same binding energy as the zero-momentum state.

After the quasiparticle energy, the largest change appears in the real part of
the self-energy, as shown in Fig.~\ref{fig:quasi_k}b.  This quantity increases
with $k$ and changes by about $5.5~{\rm MeV}$ when the momentum increases from
$k=0$ to $1~{\rm fm}^{-1}$.  The inset of Fig.~\ref{fig:quasi_k}b shows the
momentum dependence of
${\rm Im}\,\Sigma_\Lambda^{R}(k,E_{\rm qp})$.  In contrast to the real part,
the imaginary part becomes more negative as $k$ increases, although its change
is much smaller.  More explicitly, while
${\rm Im}\,\Sigma_\Lambda^{R}(0,E_{\rm qp})=-0.0115~{\rm MeV}$, it reaches
$-0.0191~{\rm MeV}$ at $k=1~{\rm fm}^{-1}$.  This means that the damping of the
$\Lambda$ quasiparticle slightly increases at larger momentum, but the
imaginary part remains very small.  Therefore, the finite-momentum
quasiparticle is still narrow and well defined.

Similarly, as shown in Figs.~\ref{fig:quasi_k}c and
\ref{fig:quasi_k}d, the momentum dependence of the quasiparticle residue and
damping width, calculated from Eqs.~(\ref{eq:Z_factor}) and
(\ref{eq:Gamma_width}), is weak.  The residue $Z(k)$ decreases with increasing
$k$, while the damping width $\Gamma(k)$ increases.  From $k=0$ to
$1~{\rm fm}^{-1}$, $Z(k)$ changes only by about $0.007$, whereas
$\Gamma(k)$ changes by about $0.015~{\rm MeV}$.  This small variation shows
that the spectral strength remains concentrated near the quasiparticle pole
over the whole momentum interval considered.  Thus, although the binding energy
of the $\Lambda$ quasiparticle changes strongly with momentum, its
quasiparticle character remains robust.

\subsubsection{$\Lambda$ spectral function}

\begin{figure}
    \centering
    \includegraphics[width=0.8\textwidth]{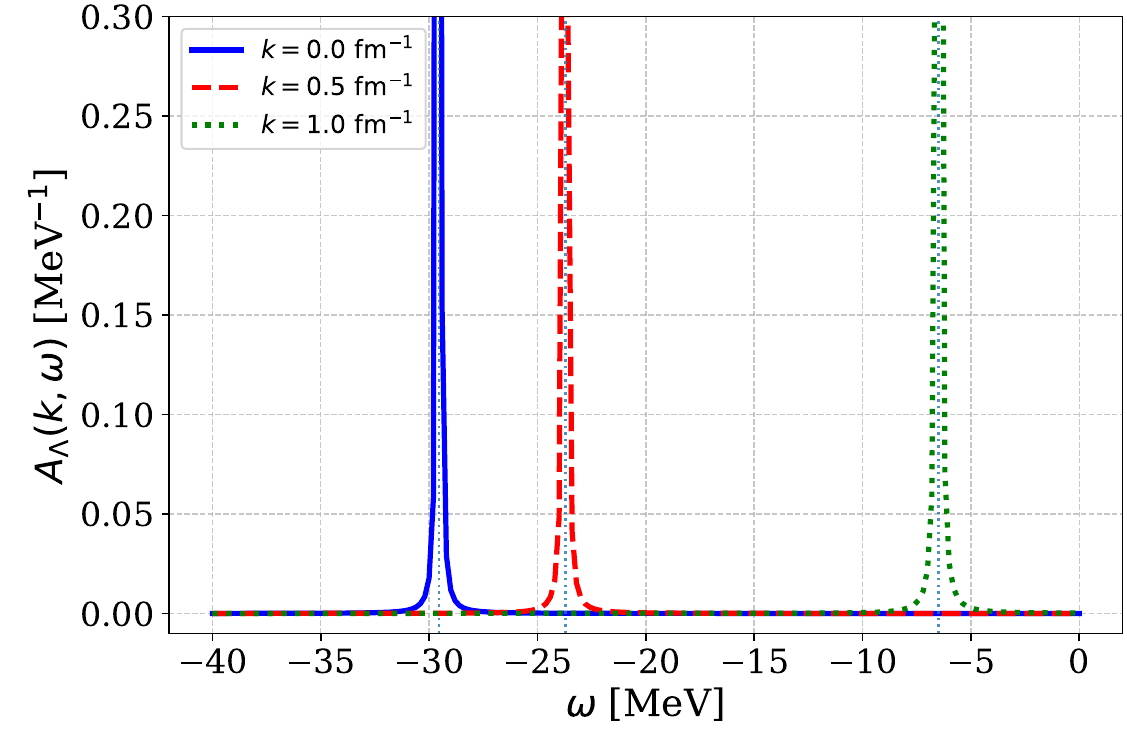}
    \caption{$\Lambda$ spectral function $A_\Lambda(k,\omega)$ for $k=0$, $0.5$, and $1~{\rm fm}^{-1}$.
    }
    \label{fig:A_Lambda_k}
\end{figure}

The calculation of the real and imaginary parts of the retarded self-energy
allows us to determine the $\Lambda$ spectral function from
Eq.~(\ref{eq:spectral_function}).  The results are shown in
Fig.~\ref{fig:A_Lambda_k} as functions of $\omega$ for
$k=0$, $0.5$, and $1~{\rm fm}^{-1}$.  The spectral function displays sharp
peaks at the corresponding quasiparticle energies,
$
E_{\rm qp}(0,\rho_{\rm sat})=-29.55~{\rm MeV},
E_{\rm qp}(0.5,\rho_{\rm sat})=-23.71~{\rm MeV},
E_{\rm qp}(1,\rho_{\rm sat})=-6.48~{\rm MeV}.
$
This shows that, at each momentum, most of the $\Lambda$ single-particle
strength is concentrated near the quasiparticle pole.  Therefore, the solution
of the quasiparticle equation is not a broad or unstable structure.  It
corresponds to a narrow and well-defined $\Lambda$ quasiparticle in symmetric
nuclear matter.

As the momentum increases, the peak moves toward larger energies, in agreement
with the finite-momentum quasiparticle dispersion discussed above.  At the same
time, the peaks remain narrow, which is consistent with the large
quasiparticle residue and the small damping width.  Thus, the spectral
function confirms that the single $\Lambda$ impurity keeps a strong
quasiparticle character in the nucleonic medium over the whole momentum
interval considered.

\subsubsection{$\Lambda$ Effective mass}

The finite-momentum quasiparticle energies shown in
Figs.~\ref{fig:Eqp} and \ref{fig:quasi_k}a allow us to determine the curvature
coefficient $c$ in the low-momentum region by using Eq.~(\ref{eq:low_k_fit}).
Then, using Eq.~(\ref{eq:mstar_fit}), we extract the effective mass of the
$\Lambda$ hyperon.  Let us note that calculations with Nijmegen
hyperon--nucleon potentials for single $\Lambda$ systems give
$m_\Lambda^*/m_\Lambda$ values between about $0.67$ and $0.81$ at saturation
density~\cite{Vidana2001}.  Therefore, the effective mass provides an important
test of whether our regulated $N\Lambda$ interaction produces a realistic
momentum dependence of the $\Lambda$ quasiparticle spectrum.

\begin{figure}
    \centering
    \includegraphics[width=0.8\textwidth]{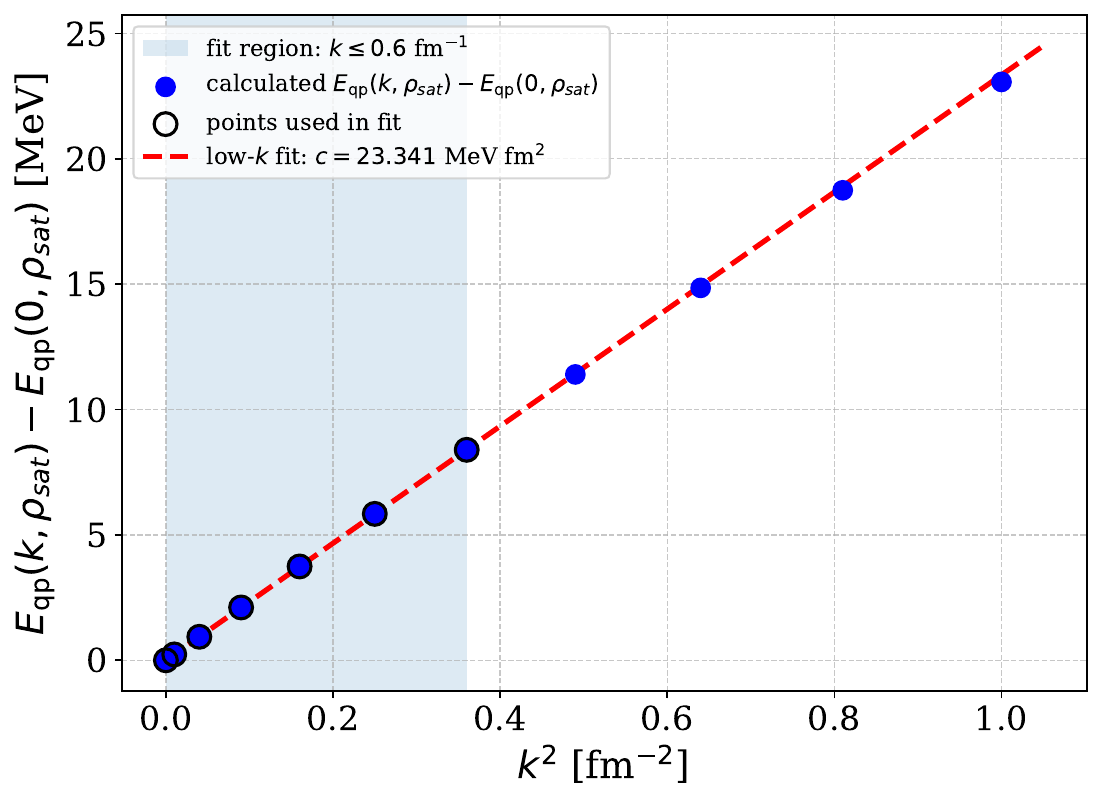}
    \caption{Low-momentum extraction of the $\Lambda$ effective mass.
    }
    \label{fig:effective_mass}
\end{figure}

For this purpose, as discussed above, we first determine the coefficient $c$
from Eq.~(\ref{eq:low_k_fit}).  The result of the low-momentum fit is shown in
Fig.~\ref{fig:effective_mass}, where we use the fitting interval
$k\leq 0.6~{\rm fm}^{-1}$.  As can be seen, the quasiparticle dispersion is
well described by a linear dependence on $k^2$ in this region, and we obtain
$
c=23.341~{\rm MeV\,fm^2}.
$
Substituting this value into Eq.~(\ref{eq:mstar_fit}) gives
\begin{equation}
\frac{m_\Lambda^*}{m_\Lambda}=0.747 .
\end{equation}
This result shows that the $\Lambda$ quasiparticle dispersion is clearly
renormalized by the nuclear medium.  The effective mass is smaller than the
free $\Lambda$ mass, which means that the quasiparticle energy increases with
momentum faster than the free-particle kinetic energy near $k=0$.  At the same
time, the value obtained here lies inside the range reported in
Ref.~\cite{Vidana2001}.  Thus, the regulated $N\Lambda$ interaction used in
the present work gives a realistic effective-mass scale for a single $\Lambda$
in nuclear matter.

Let us also note that the extracted value is stable against reasonable changes
of the fitting interval.  As shown in Table~\ref{tab:effective_mass}, choosing
$k_{\rm fit}^{\rm max}=0.2~{\rm fm}^{-1}$ changes
$m_\Lambda^*/m_\Lambda$ by only about $0.002$ compared with the result obtained
for $k_{\rm fit}^{\rm max}=0.6~{\rm fm}^{-1}$.  This weak dependence on the
fit interval indicates that the quadratic low-momentum approximation is stable
and that the choice $k_{\rm fit}^{\rm max}=0.6~{\rm fm}^{-1}$ is reliable for
extracting the effective mass.
 
\begin{table}[t]\label{tab:effective_mass}
\centering
\caption{
Sensitivity of the extracted $\Lambda$ effective mass to the low-momentum
fitting interval.
}
\begin{tabular}{c c c}
\hline\hline
$k_{\rm fit}^{\max}$ (fm$^{-1}$) & $c$ (MeV fm$^2$) &
$m_\Lambda^*/m_\Lambda$ \\
\hline
0.2 & 23.446 & 0.744 \\
0.4 & 23.402 & 0.746 \\
0.6 & 23.341 & 0.747 \\
\hline\hline
\end{tabular}
\end{table}

\section{Conclusion}\label{sec:conc}  

We have studied the quasiparticle properties of a single $\Lambda$ hyperon propagating through symmetric nuclear matter. The calculation was performed in the impurity limit, where $\rho_\Lambda=0$ and $\rho_n=\rho_p=\rho_{\rm sat}/2$. In this limit, the nucleons form the occupied Fermi sea, while the $\Lambda$ hyperon is treated as a distinct quasiparticle probing the nucleonic medium. 
The elementary input of the calculation was a non-local regulated low-momentum $N\Lambda$ contact interaction. The potential contains a leading-order constant term and a next-to-leading-order derivative correction. For the ${}^1S_0$ and ${}^3S_1$ channels, the corresponding coupling constants were fixed by matching the vacuum on-shell $T$ matrix to the scattering length and effective range obtained from modern next-to-next-to-leading-order chiral effective field theory. For spin-saturated symmetric nuclear matter, the interaction was then taken as the statistical spin average of the singlet and triplet channels. This gives a compact effective-range-constrained $N\Lambda$ interaction that keeps the leading low-energy attraction and the momentum dependence needed for the quasiparticle dispersion.

Using this interaction, we calculated the retarded $\Lambda$ self-energy from the in-medium $N\Lambda$ ladder $T$ matrix. In this way, the self-energy contains both the static Born contribution and the dynamical contribution generated by repeated in-medium $N\Lambda$ scattering. At saturation density, the zero-momentum quasiparticle pole was found at
$E_{\rm qp}(0,\rho_{\rm sat})=-29.55~{\rm MeV}$, in good agreement with the empirical depth of the single $\Lambda$ potential in nuclear matter. The decomposition of the self-energy showed that the Born contribution gives $\Sigma_\Lambda^{\rm Born}(0)=-26.36~{\rm MeV}$, while the correlation part gives ${\rm Re}\,\Sigma_\Lambda^{\rm corr,R}(0,E_{\rm qp})=-3.19~{\rm MeV}$. Therefore, although the dominant attraction comes from the static Born term, the empirical binding scale is reached only after the dynamical correlation contribution from repeated $N\Lambda$ scattering is included.

The zero-momentum $\Lambda$ quasiparticle was found to be narrow and well defined. We obtained a large quasiparticle residue, $Z(0)=0.98$, and a small damping width, $\Gamma(0)=0.023~{\rm MeV}$. The corresponding spectral function shows a sharp peak at the quasiparticle energy. These results demonstrate that most of the $\Lambda$ single-particle strength remains concentrated near the pole. Thus, the attractive solution of the quasiparticle equation is not a broad or unstable structure, but a robust $\Lambda$ quasiparticle in symmetric nuclear matter.

We also extended the calculation to finite momenta. The quasiparticle energy becomes less bound as the momentum increases, changing from $E_{\rm qp}(0,\rho_{\rm sat})=-29.55~{\rm MeV}$ at $k=0$ to $E_{\rm qp}(1~{\rm fm}^{-1},\rho_{\rm sat})=-6.49~{\rm MeV}$. In contrast, the quasiparticle residue and damping width change only weakly over the same momentum interval. This shows that the main momentum dependence appears in the quasiparticle dispersion, while the pole strength remains large and the damping remains small. The spectral function at finite momenta confirms the same picture: the peak moves to higher energy as $k$ increases, but it remains narrow and clearly identifiable.

From the low-momentum quasiparticle dispersion, we extracted the effective mass of the $\Lambda$ hyperon. Using the fit of $E_{\rm qp}(k,\rho_N)-E_{\rm qp}(0,\rho_N)$ as a function of $k^2$, we obtained $m_\Lambda^*/m_\Lambda=0.747$ at saturation density. This value lies in the range found in Brueckner calculations with Nijmegen hyperon--nucleon potentials. Therefore, the present regulated $N\Lambda$ interaction reproduces not only the empirical attraction of the $\Lambda$ impurity, but also a realistic momentum dependence of the quasiparticle spectrum.

The main result of this work is that a compact effective-range-constrained $N\Lambda$ interaction, combined with an in-medium ladder self-energy, gives a realistic and transparent description of a single $\Lambda$ impurity in symmetric nuclear matter. The calculation reproduces the empirical single $\Lambda$ potential depth, predicts a narrow quasiparticle with large pole strength, and gives an effective mass consistent with microscopic hypernuclear-matter calculations. This provides a useful two-body correlated baseline for future extensions in which finite $\Lambda$ density, $\Lambda\Lambda$ interactions, coupled $\Lambda N$--$\Sigma N$ channels, three-baryon forces, and beta-equilibrated hyperonic matter will be included.

\appendix 
\section{Mapping of $N\Lambda$ Effective-Range Parameters to Contact Couplings} \label{sec:A1}
The constants $g_{0,N\Lambda}^{S}$ and $g_{2,N\Lambda}^{S}$ are not independent
physical observables. They are cutoff-dependent low-energy constants of the
regulated $N\Lambda$ interaction. The physical input is the scattering length $a_{N\Lambda}^{S}$
and effective range $r_{N\Lambda}^{S}$ in each $S$-wave channel.
After choosing the cutoff $\Lambda_{\rm cut}$, the constants
$g_{0,N\Lambda}^{S}$ and $g_{2,N\Lambda}^{S}$ are fixed so that the vacuum
on-shell $T$ matrix reproduces the effective-range expansion. In this work we
keep the two dominant $N\Lambda$ $S$-wave channels:  ${}^1S_0$ and $ {}^3S_1$.
For each spin channel $S=0,1$, we solve the vacuum two-body
Lippmann--Schwinger equation
\begin{eqnarray}
T_{N\Lambda}^{S}(p',p;E)
&=&
V_{N\Lambda}^{S}(p',p)
+
\int\!\frac{d^3q}{(2\pi)^3}\,
V_{N\Lambda}^{S}(p',q)
\frac{1}{E-\frac{q^2}{2\mu_{N\Lambda}}+i0^+}
T_{N\Lambda}^{S}(q,p;E),\nonumber\\
\label{eq:LS_vacuum_NL}
\end{eqnarray}
where $\mu_{N\Lambda} = m_Nm_\Lambda / \left(m_N+m_\Lambda\right)$
is the reduced mass. The on-shell energy is $E=\hbar^2k^2 / \left(2\mu_{N\Lambda}\right)$.
The on-shell $T$ matrix is related to the scattering amplitude by \cite{Fetter2003,Taylor2012}
\begin{equation}
f_{N\Lambda}^{S}(k)
=
-\frac{\mu_{N\Lambda}}{2\pi}\,
T_{N\Lambda}^{S}(k,k;E).
\label{eq:f_T_NL}
\end{equation}
Therefore,
\begin{equation}
k\cot\delta_{N\Lambda}^{S}(k)
=
-\frac{2\pi}{\mu_{N\Lambda}}
\operatorname{Re}
\left[
\frac{1}{T_{N\Lambda}^{S}(k,k;E)}
\right].
\label{eq:kcot_T_NL}
\end{equation}
The two constants $g_{0,N\Lambda}^{S}$ and $g_{2,N\Lambda}^{S}$ are then fixed
by requiring
\begin{equation}
k\cot\delta_{N\Lambda}^{S}(k)\Big|_{k\rightarrow 0}
=
-\frac{1}{a_{N\Lambda}^{S}},
\label{eq:match_a_NL}
\end{equation}
and
\begin{equation}
\left.
\frac{d}{dk^2}
k\cot\delta_{N\Lambda}^{S}(k)
\right|_{k=0}
=
\frac{1}{2}r_{N\Lambda}^{S}.
\label{eq:match_r_NL}
\end{equation}
In this way, the regulated interaction used in the many-body calculation is
connected directly to the low-energy $N\Lambda$ scattering observables. The
resulting values of $g_{0,N\Lambda}^{S}$ and $g_{2,N\Lambda}^{S}$ for
$\Lambda_{\rm cut}=500~{\rm MeV}$ are listed in Table~\ref{tab:scatt_input}.

\bibliographystyle{elsarticle-num}
\bibliography{bib}
\end{document}